\documentclass[%
 aip,
 jmp,%
 amsmath,amssymb,
 reprint,%
]{revtex4-1}
 
\usepackage{graphicx}
\usepackage{fullpage}
\usepackage{dcolumn}
\usepackage{bm}

\begin{document}
 
\preprint{AIP/123-QED}

\title{Predicting the outcome of roulette}
 
\author{Michael Small}
\email{michael.small@uwa.edu.au}
\affiliation{School of Mathematics and Statistics, The University of Western Australia}
\affiliation{Department of Electronic and Information Engineering\\ 
Hong Kong Polytechnic University, Hong Kong.}
\author{Chi Kong Tse}
\affiliation{Department of Electronic and Information Engineering\\ 
Hong Kong Polytechnic University, Hong Kong.}

\date{\today}

\begin{abstract}
There have been several popular reports of various groups exploiting the deterministic nature of the game of roulette for profit. Moreover, through its history the inherent determinism in the game of roulette has attracted the attention of many luminaries of chaos theory. In this paper we provide a short review of that history and then set out to determine to what extent that determinism can really be exploited for profit. To do this, we provide a very simple model for the motion of a roulette wheel and ball and demonstrate that knowledge of initial position, velocity and acceleration is sufficient to predict the outcome with adequate certainty to achieve a positive expected return. We describe two physically realisable systems to obtain this knowledge both incognito and {\em in situ}. The first system relies only on a mechanical count of rotation of the ball and the wheel to measure the relevant parameters. By applying this techniques to a standard casino-grade European roulette wheel we demonstrate an expected return of at least  $18\%$, well above the $-2.7\%$ expected of a random bet. With a more sophisticated, albeit more intrusive, system (mounting a digital camera above the wheel) we demonstrate a range of systematic and statistically significant biases which can be exploited to provide an improved guess of the outcome. Finally, our analysis demonstrates that even a very slight slant in the roulette table leads to a very pronounced bias which could be further exploited to substantially enhance returns.
\end{abstract}

\maketitle

\begin{center}
``No one can possibly win at roulette unless he\\
 steals money from the table when the croupier\\ 
 isn't looking''
 (Attributed to Albert Einstein in \cite{tB90})
\end{center}

\begin{quotation}
Among the various gaming systems, both current and historical, roulette is uniquely deterministic. Relatively simple laws of motion allow one, in principle, to forecast the path of the ball on the roulette wheel and to its final destination. Perhaps because of this appealing deterministic nature, many notable figures from the early development of chaos theory have leant their hand to exploiting this determinism and undermining the presumed randomness of the outcome. In this paper we aim only to establish whether the determinism in this system really can be profitably exploited. We find that this is definitely possible and propose several systems which could be used to gain an edge over the house in a game of roulette. While none of these systems are optimal, they all demonstrate positive expected return.
\end{quotation}

\section{A history of roulette}

The game of roulette has a long, glamorous, inglorious history, and has been connected with several notable men of science. The origin of the game has been attributed \cite{rE67}, perhaps erroneously \cite{tB90}, to the mathematician Blaise Pascal \cite{eB37}. Despite the roulette wheel becoming a staple of probability theory, the alleged motivation for Pascal's interest in the device was not solely to torment undergraduate students, but rather as part of a vain search for perpetual motion. Alternative stories have attributed the origin of the game to the ancient Chinese, a French monk or an Italian mathematician\footnote{The Italian mathematician, confusingly, was named Don Pasquale \cite{rE67}, a surname phonetically similar to Pascal. Moreover, as Don Pasquale is also the name of a 19th century opera buff, this attribution is possibly fanciful.}$\;$ \cite{rE67}. In any case, the device was introduced to Parisian gamblers in the mid-eighteenth century to provide a fairer game than those currently in circulation. By the turn of the century, the game was popular and wide-spread. Its popularity bolstered by its apparent randomness and inherent (perceived) honesty. 

\begin{figure*}
\begin{center}
\begin{tabular}{cc}
\includegraphics[width=0.45\textwidth]{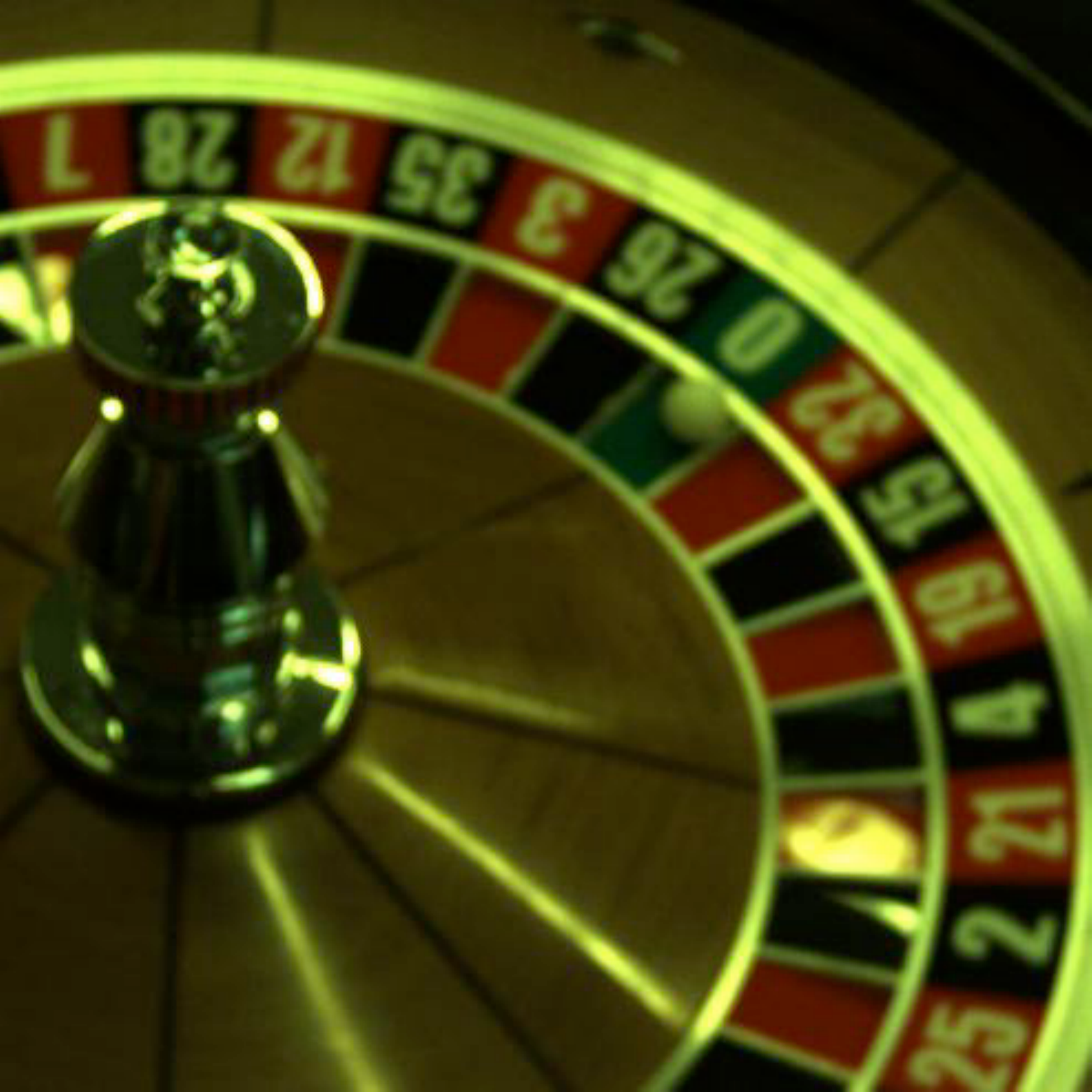}
&
\includegraphics[width=0.45\textwidth]{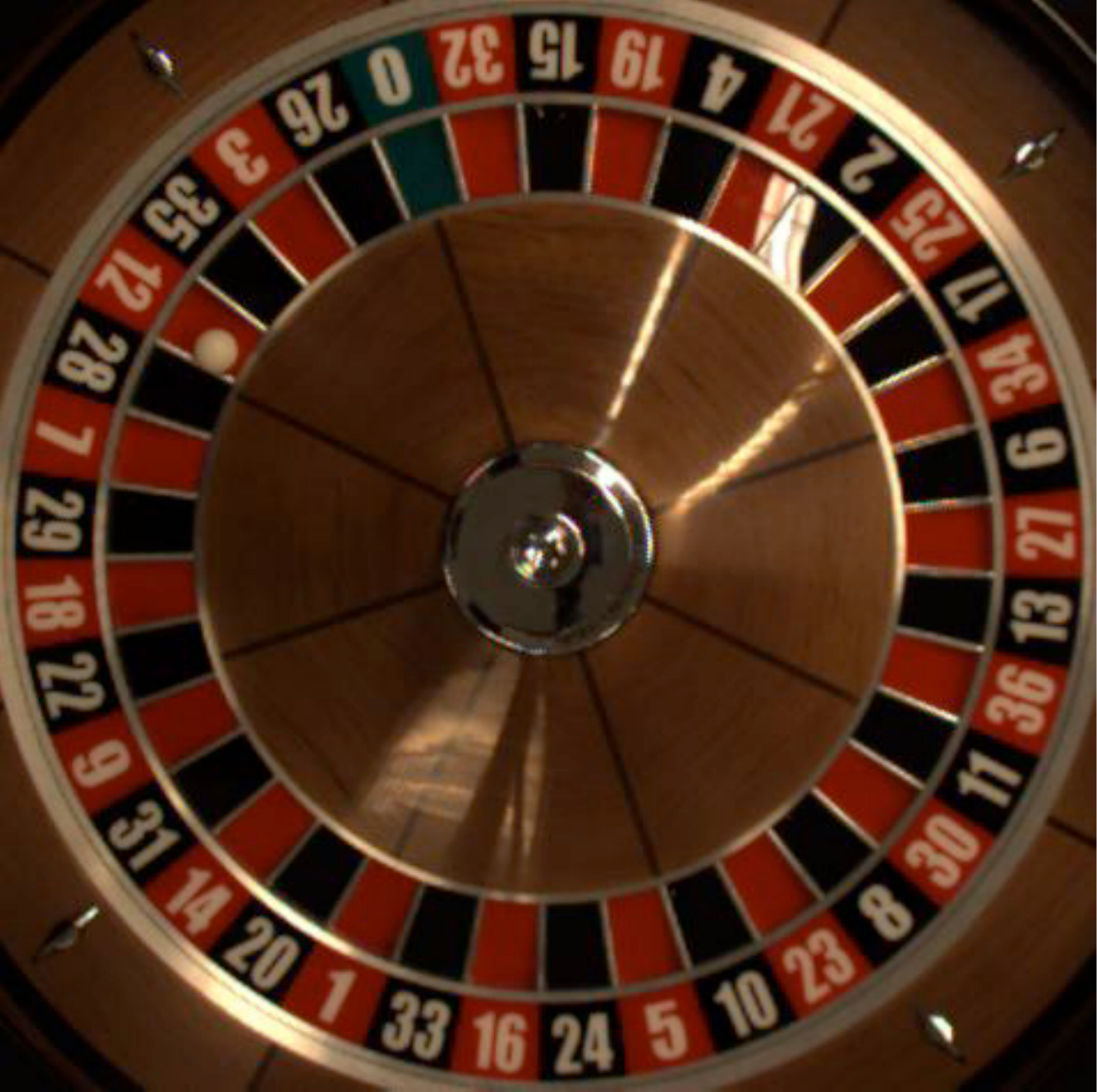}
\end{tabular}
\end{center}
\caption{{\bf The European roulette wheel.} In the left panel one can see a portion of the rotating roulette wheel and surrounding fixed track. The ball has come to rest in the green 0 pocket. Although the motion of the wheel and the ball (in the outer track) are simple and linear, one can see the addition  of several  metal deflectors on the {\em stator} (that is the fixed frame on which the rotating wheel sits).
The sharp {\em frets} between pockets also introduce strong nonlinearity as the ball slows and bounces between pockets. The  panel on the right  depicts the arrangement of the number 0 to 36 and the colouring red and black.}
\label{wheel}
\end{figure*}

The game of roulette consists of a heavy wheel, machined and balanced to have very low friction and designed to spin for a relatively long time with a slowly decaying angular velocity. The wheel is spun in one direction, while a small ball is spun in the opposite direction on the rim of a fixed circularly inclined surface surrounding and abutting the wheel. As the ball loses momentum it drops toward the wheel and eventually will come to rest in one of 37 numbered pockets arranged around the outer edge of the spinning wheel.  Various wagers can be made on which pocket, or group of pockets, the ball will eventually fall into. It is accepted practise that, on a successful wager on a single pocket, the casino will pay 35 to 1. Thus the expected return from a single wager on a fair wheel is $(35+1)\times\frac{1}{37}+(-1)\approx -2.7\%$ \cite{fD72}. In the long-run, the house will, naturally, win. In the eighteenth century the game was fair and consisted of only 36 pockets. Conversely, an American roulette wheel is even less fair and consists of 38 pockets. We consider the European, 37 pocket, version as this is of more immediate interest to us \cite{bO79}. Figure \ref{wheel} illustrates the general structure, as well as the layout of pockets, on a standard European roulette wheel.

Despite many proposed ``systems'' there are only two profitable ways to play roulette\footnote{Three, if one has sufficient finances to assume the role of the house.}. One can either exploit an unbalanced wheel, or one can exploit the inherently deterministic nature of the spin of both ball and wheel. Casinos will do their utmost to avoid the first type of exploit. The second exploit is possible because placing wagers on the outcome is traditionally permitted until some time after the ball and wheel are in motion. That is, one has an opportunity to observe the motion of both the ball and the wheel before placing a wager.

The archetypal tale of the first type of exploit is that of a man by the name of Jagger (various sources refer to him as either William Jaggers or Joseph Jagger, or some permutation of these). Jagger, an English mechanic and amateur mathematician, observed that slight mechanical imperfection in a roulette wheel could afford sufficient edge to provide for profitable play. According to one incarnation of the tale, in 1873 he embarked for the casino of Monte Carlo with six hired assistants. Once there, he carefully logged the outcome of each spin of each of six roulette tables over a period of five weeks \cite{cK25}. Analysis of the data revealed that for each wheel there was a unique but systematic bias. Exploiting these weaknesses he gambled profitably for a week before the casino management shuffled the wheels between tables. This bought his winning streak to a sudden halt. However, he soon noted various distinguishing features of the individual wheels and was able to follow them between tables, again winning consistently. Eventually the casino resorted to redistributing the individual partitions between pockets. A popular account, published in 1925, claims he eventually came away with winnings of $\pounds 65,000$ \cite{cK25}. The success of this endeavour is one possible inspiration for the musical hall song ``The Man Who Broke the Bank at Monte Carlo'' although this is strongly disputed \cite{cK25}. 

Similar feats have been repeated elsewhere. The noted statistician Karl Pearson provided a statistical analysis of roulette data, and found it to exhibit substantial systematic bias. However, it appears that his analysis was based on flawed data from unscrupulous scribes \cite{eT69} (apparently he had hired rather lazy journalists to collect the data). 

In 1947 irregularities were found, and exploited, by two students, Albert Hibbs and Roy Walford, from Chicago University \cite{life47}\;\footnote{Alternatively, and apparently erroneously, reported to be from  Californian Institute of Technology in \cite{rE67}}. Following this line of attack, S.N. Ethier provides a statistical framework by which one can test for irregularities in the observed outcome of a roulette wheel \cite{sE82}. A similar weakness had also been reported in {\em Time} magazine in 1951. In this case, the report described various syndicates of gamblers exploiting determinism in the roulette wheel in the Argentinean casino Mar del Plata during 1948 \cite{time51}. The participants were colourfully described as a Nazi sailor and  various ``fruit hucksters, waiters and farmers'' \cite{time51}. 

The second type of exploit is more physical (that is, deterministic) than purely statistical and has consequently attracted the attention of several mathematicians, physicists and engineers. One of the first\footnote{The first, to the best of our knowledge.} was Henri Poincar\'e \cite{eB37} in his seminal work {\em Science and Method} \cite{hP14}. While ruminating on the nature of chance, and that a small change in initial condition can lead to a large change in effect, Poincar\'e illustrated his thinking with the example of a roulette wheel (albeit a slightly different design from the modern version). He observed that a tiny change in initial velocity would change the final resting place of the wheel (in his model there was no ball) such that the wager on an either black or red  (as in a modern wheel, the black and red pockets alternate) would correspondingly win or lose. He concluded by arguing that this determinism was not important in the game of roulette as the variation in initial force was tiny, and for any {\em continuous} distribution of initial velocities, the result would be the same: effectively random, with equal probability. He was not concerned with the individual pockets, and he further assumed that the variation in initial velocity required to predict the outcome would be immeasurable. It is while describing the game of roulette that Poincar\'e introduces the concept of sensitivity to initial conditions, which is now a cornerstone of modern chaos theory \cite{jC86b}.

A general procedure for predicting the outcome of a roulette spin, and an assessment of its utility was described by Edward Thorp in a 1969 publication for the {\em Review of the International Statistical Institute} \cite{eT69}. In that paper, Thorp describes the two basic methods of prediction. He observes (as others have done later) that by minimising systematic bias in the wheel, the casinos achieve a mechanical perfection that can then be exploited using deterministic prediction schemes. He describes two deterministic prediction schemes (or rather two variants on the same scheme). If the roulette wheel is not perfectly level (a tilt of $0.2^{\circ}$ was apparently sufficient --- we verified that this is indeed more than sufficient) then there effectively is a large region of the frame from which the ball will not fall onto  the spinning wheel. By studying Las Vegas wheels he observes this condition is meet in approximately one third of wheels. He claims that in such cases it is possible to garner a expectation of $+15\%$, which increased to $+44\%$ with the aid of a `pocket-sized' computer. Some time later, Thorp revealed that his collaborator in this endeavour was Claude Shannon \cite{eT85}, the founding father of information theory \cite{cS48}. 

In his 1967 book \cite{rE67} the mathematician Richard A. Epstein describes his earlier (undated) experiments with a private roulette wheel. By measuring the angular velocity of the ball relative to the wheel he was able to predict correctly the half of the wheel into which the ball would fall. Importantly, he noted that the initial velocity (momentum) of the ball was not critical. Moreover, the problem is simply one of predicting when the ball will leave the outer (fixed rim) as this will always occur at a fixed velocity. However, a lack of sufficient computing resources meant that his experiments were not done in real time, and certainly not attempted within a casino. 

Subsequent to, and inspired by, the work of Thorp and Shannon, another widely described attempt to beat the casinos of Las Vegas was made in 1977-1978 by Doyne Farmer, Norman Packard and colleagues \cite {tB90}. It is supposed that Thorp's 1969 paper had let the cat out of the bag regarding profitable betting on roulette. However, despite the assertions of Bass\cite{tB90}, Thorp's paper \cite{eT69} is not mathematically detailed (there is in fact no equations given in the description of roulette). Thorp is sufficiently detailed to leave the reader in no doubt that the scheme could work, but also vague enough so that one could not replicate his effort without considerable knowledge and skill. Farmer, Packard and colleagues implemented the system on a 6502 microprocessor hidden in a shoe, and proceeded to apply their method to the various casinos of the Las Vegas Strip. The exploits of this group are described in detail in Bass\cite{tB90}. The same group of physicists went on to apply their skills to the study of chaotic dynamical systems \cite{nP80} and also for profitable trading on the financial markets  \cite{tB99}.  In Farmer and Sidorowich's landmark paper on predicting chaotic time series \cite{jF87} the authors attribute the inspiration for that work  to their earlier efforts to beat the game of roulette. 

Less exalted individuals have also been employing similar schemes, in some cases fairly recently. In 2004, the BBC carried the report of three gamblers (described only as ``a Hungarian woman and two Serbian men'' \cite{bbc04-1}) arrested by police after winning $\pounds 1,300,000$ at the Ritz Casino in London. The trio had apparently been using a laser scanner and their mobile phones to predict the likely resting place of the ball. Happily, for the trio but not the casino, they were judged to have broken no laws and allowed to keep their winnings \cite{bbc04-2}. The scheme we describe in Section \ref{model} and implement in Section \ref{predict} is certainly compatible with the equipment and results reported in this case. In Section \ref{exploit} we conclude with some remarks concerning the practicality of applying these methods in a modern casino, and what steps a casino could take (or perhaps have taken) to circumvent these exploits. A preliminary version of these results was presented at a conference in Macao \cite{roulette}. An independent and much more detailed model of dynamics of the roulette wheel is discussed in Strzalko {\em et al.} \cite{jS09}. Since our  preliminary publication\cite{roulette}, private communication with several individuals indicates that these methods have now progressed to the point of at least four instances of independent {\em in situ} field trials.

\section{A model for roulette}
\label{model}

\begin{figure}
\begin{center}
\includegraphics[width=0.45\textwidth]{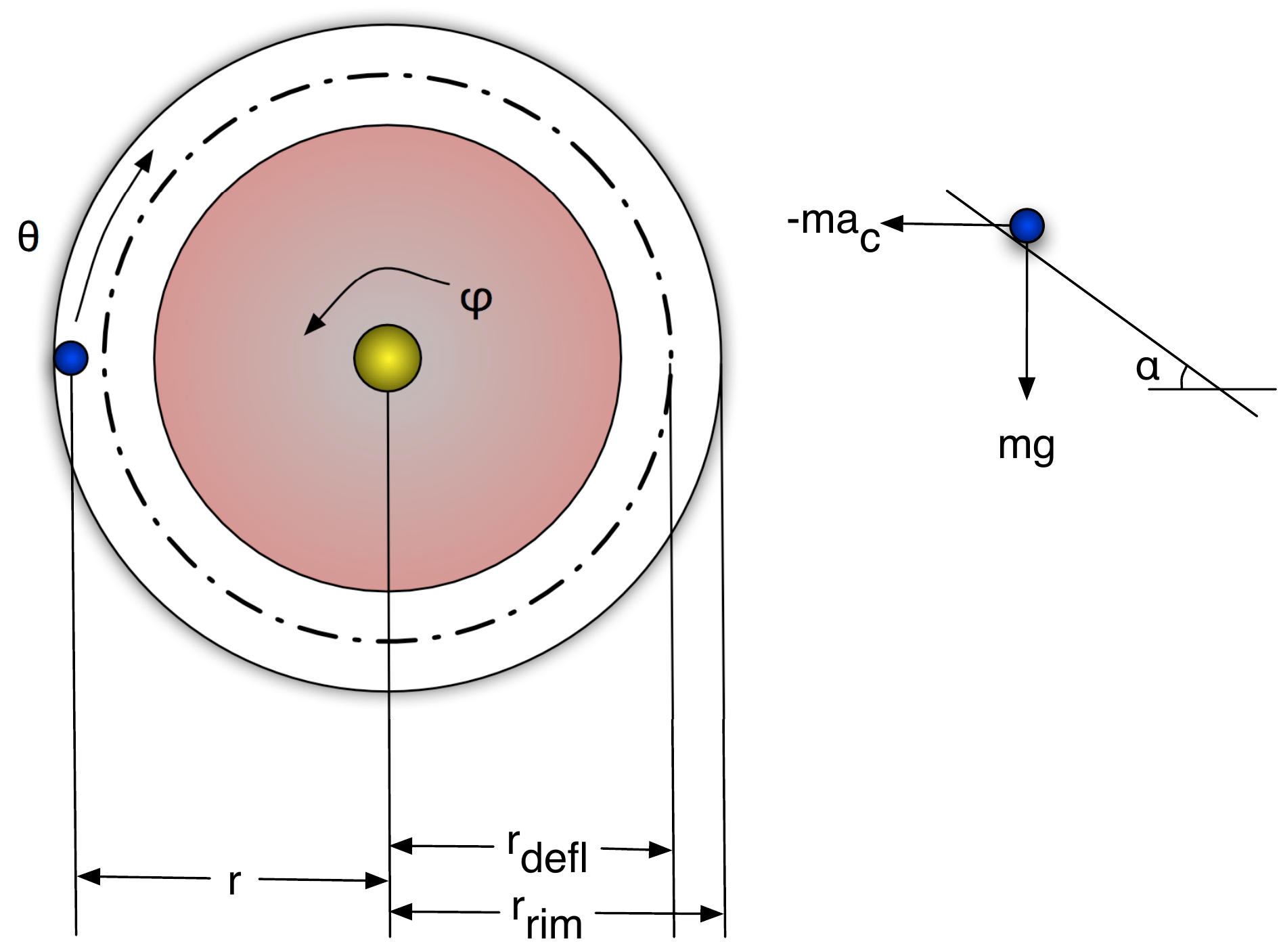}
\end{center}
\caption{{\bf The dynamic model of ball and wheel.} On the left we show a top view of the roulette wheel (shaded region) and the stator (outer circles). The ball is moving on the stator with instantaneous position $(r,\theta)$ while the wheel is rotating with angular velocity $\dot\varphi$ (Note that the direction of the arrows here are for illustration only, the analysis in the text assume the same convention, clockwise positive, for both ball and wheel). The deflectors on the stator are modelled as a circle, concentric with the wheel, of radius $r_{\rm defl}$. On the right we show a cross section and examination of the forces acting on the ball in the incline plane of the stator. The angle $\alpha$ is the incline of the stator, $m$ is the mass of the ball, $a_{\rm c}$ is the radial acceleration of the ball, and $g$ is gravity.}
\label{forces}
\end{figure} 

We now describe our basic model of the motion of the roulette wheel and ball. Let $(r,\theta)$ denote the position of the ball in polar co-ordinates, and let $\varphi$ denote the angular position of the wheel (say, the angular position of the centre of the green 0 pocket). We will model the ball as a single point and so let $r_{\rm rim}$ be the farthest radial position of that point (i.e. the radial position of the centre of the ball when the ball is spinning with high velocity in the rim of the wheel). Similarly, let $r_{\rm defl}$ be the radial distance to the location of the metal deflectors on the stator. For now, we will assume that $\frac{dr_{\rm defl}}{d\theta}=0$ (that is, there are deflectors evenly distributed around the stator at constant radius $r_{\rm defl}<r$). The extension to the more precise case is obvious, but, as we will see, not necessary. Moreover, it is messy. Finally, we suppose that the incline of the stator to the horizontal is a constant $\alpha$. This situation, together with a balance of forces is depicted in figure \ref{forces}. We will first consider the ideal case of a level table, and then in section \ref{unlevel} show how this condition is in fact  critical.

\subsection{Level table}
\label{level}

For a given initial motion of ball $(r,\theta,\dot\theta,\ddot\theta)_{t=0}$ and wheel $(\varphi,\dot\varphi,\ddot\varphi)_{t=0}$ our aim is to determine the time $t_{\rm defl}$ at which $r=r_{\rm defl}$. After launch the motion of the ball will pass through two distinct states which we further divide into four cases: (i) with sufficient momentum it will remain in the rim, constrained by the fixed edge of the stator; (ii) at some point the momentum drops and the ball leaves the rim; (iii) the ball will gradually loose momentum while travelling on the stator as $\dot{\theta}$ drops, so will $r$; and (iv) eventually $r=r_{\rm defl}$ at some time $t_{\rm defl}$. At time $t=t_{\rm defl}$ we assume that the ball hits a deflector on the stator and drops onto the (still spinning) wheel. Of course, the deflectors are discrete and located only at specific points around the edge of the wheel. While it is possible, and fairly straightforward to incorporate the exact position (and more importantly, the orientation) of each deflector, we have not done this. Instead, we model the deflectors at a constant radial distance around the entire rim. The exact position of the wheel when the ball reaches the deflectors will be random but will depend only on $\varphi(t_{\rm defl})$ --- i.e. depending on where the actual deflectors are when the ball first comes within range, the radial distance until the ball actually deflects will be uniformly distributed on the interval $[0,2\pi/N_{\rm defl}]$ where $N_{\rm defl}$ is the number of deflectors.  

\subsubsection*{(i) Ball rotates in the rim}

While traveling in the rim $r$ is constant and the ball has angular velocity $\dot\theta$. Hence, the radial acceleration of the ball is $a_{c}=\frac{v^2}{r}=\frac 1r (r\dot\theta)^2=r\dot\theta^2$ where $v$ is the speed of the ball. During this period of motion, we suppose that $r$ is constant and that $\theta$ decays only due to constant rolling friction: hence $\dot r=0$ and $\ddot\theta=\ddot\theta(0)$, a constant. This phase of motion will continue provided the centripetal force of the ball on the rim exceeds the force of gravity $ma_{\rm c}\cos\alpha>mg\sin\alpha$ ($m$ is the mass of the ball). Hence, at this stage 
\begin{eqnarray}
\label{inrim}
\dot\theta^2&>&\frac gr\tan{\alpha}.
\end{eqnarray} 

\subsubsection*{(ii) Ball leaves the rim}

Gradually the speed on the ball decays until eventually $\dot\theta^2=\frac gr\tan{\alpha}$. Given the initial acceleration $\ddot\theta(0)$, velocity $\dot\theta(0)$ and position $\theta(0)$, it is trivial to compute the time at which the ball leaves the rim, $t_{\rm rim}$ to be 
\begin{eqnarray}
\label{leaverim}
t_{\rm rim} & = &-\frac 1{\ddot\theta(0)}\left(\dot\theta(0)-\sqrt{\frac gr\tan\alpha}\right).
\end{eqnarray}
To do so, we assume that the angular acceleration is constant and so the angular velocity at any time is given by $\dot\theta(t)=\dot\theta(0)+\ddot\theta(0)t$ and substitute into equation (\ref{inrim}). That is, we are assuming that the force acting on the ball is independent of velocity --- this is a simplifying assumption for the naive model we describe here, more sophisticated alternatives are possible, but in all cases this will involve the estimation of additional parameters. The position at which the ball leaves the rim is given by
\[\left|\theta(0)+ \frac{(\frac gr\tan\alpha) -\dot\theta(0)^2}{2\ddot\theta(0)}\right|_{2\pi}\]
where $|\cdot |_{2\pi}$ denotes modulo $2\pi$.

\subsubsection*{(iii) Ball rotates freely on the stator}

After leaving the rim the ball will continue (in practise, for only a short while) to rotate freely on the stator until it eventually reaches the various deflectors at $r=r_{\rm defl}$. The angular velocity continues to be governed by 
\[\dot\theta(t)=\dot\theta(0)+\ddot\theta(0)t,\] but now that 
\[r\dot\theta^2<g\tan\alpha\] the radial position is going to gradually decrease too. The difference between the force of gravity $mg\sin\alpha$ and the (lesser) centripetal force $mr\dot\theta^2\cos\alpha$ provides inward acceleration of the ball
\begin{eqnarray}
\label{falling}
\ddot{r} &=&r\dot\theta^2\cos\alpha-g\sin\alpha.
\end{eqnarray}
Integrating (\ref{falling}) yields the position of the ball on the stator.

\subsubsection*{(iv) Ball reaches the deflectors}

Finally, we find the time $t=t_{\rm defl}$ for which $r(t)$, computed as the definite second integral of (\ref{falling}), is equal to $r_{\rm defl}$. We can then compute the instantaneous angular position of the ball $\theta(t_{\rm defl})=\theta(0)+\dot\theta(0)t_{\rm defl}+\frac 12\ddot\theta(0)t_{\rm defl}^2$ and the wheel $\varphi(t_{\rm defl})=\varphi(0)+\dot\varphi(0)t_{\rm defl}+\frac 12\ddot\varphi(0)t_{\rm defl}^2$to give the salient value 
\begin{eqnarray}
\label{rest}
\gamma & = & \left|\theta(t_{\rm defl})-\varphi(t_{\rm defl})\right|_{2\pi}
\end{eqnarray}
denoting the angular location on the wheel directly below the point at which the ball strikes a deflector. Assuming the constant distribution of deflectors around the rim, some (still to be estimated) distribution of resting place of the ball will depend only on that value $\gamma$. Note that, although we have described $(\theta,\dot\theta,\ddot\theta)_{t=0}$ and  $(\varphi,\dot\varphi,\ddot\varphi)_{t=0}$ separately, it is possible to adopt the rotating frame of reference of the wheel and treat $\theta-\varphi$ as a single variable. The analysis is equivalent, estimating the required parameters may become simpler.

We note that for a level table, each spin of the ball alters only the time spent in the rim, the ball will leave the rim of the stator with exactly the same velocity $\dot\theta$ each time. The descent from this point to the deflectors will therefore be identical. There will, in fact, be some characteristic duration which could be easily computed for a given table. Doing this would circumvent the need to integrate $(\ref{falling})$. 

\subsection{The crooked table}
\label{unlevel}

\begin{figure}
\begin{center}
\includegraphics[width=0.45\textwidth]{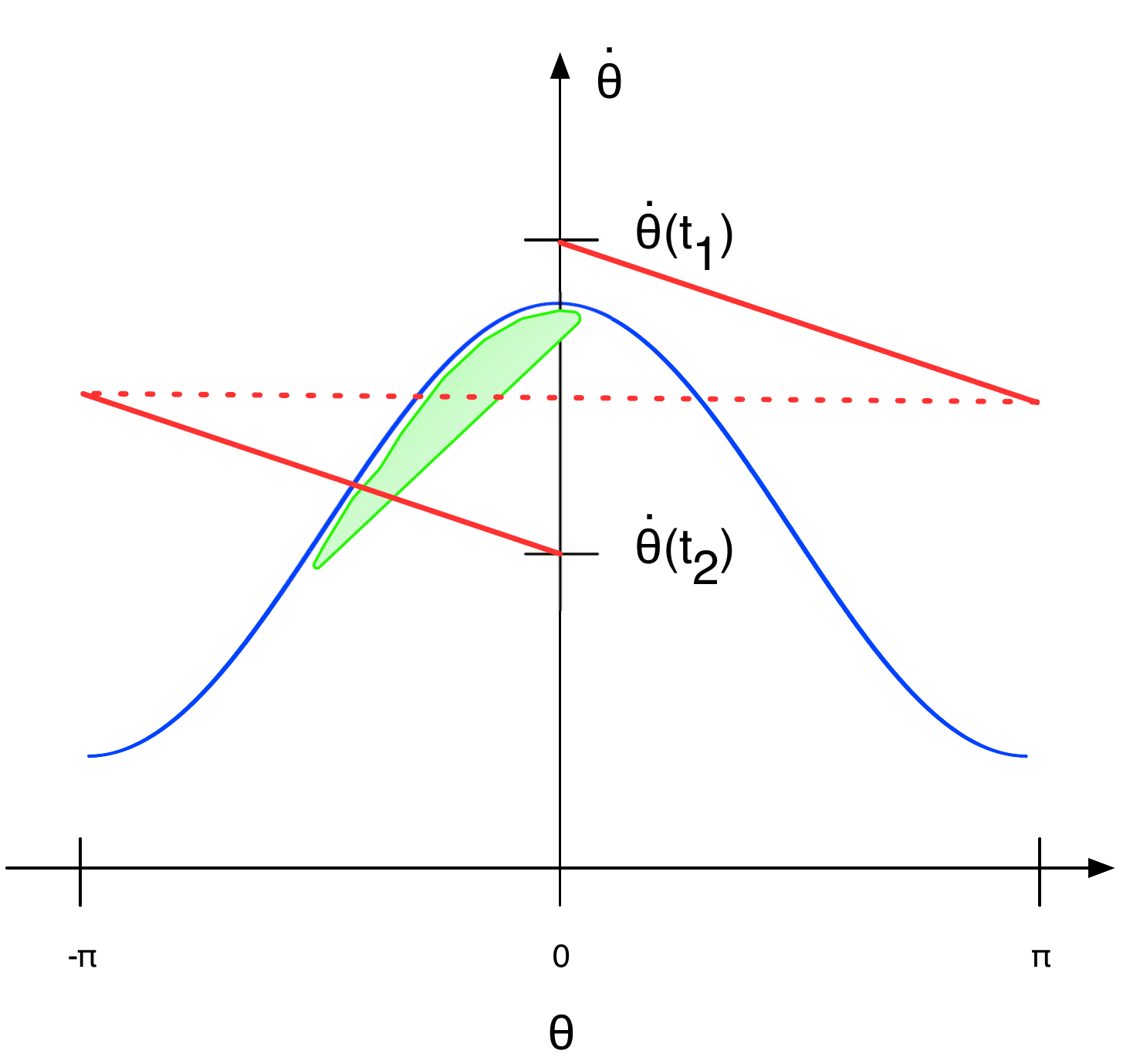}
\end{center}
\caption{{\bf The case of the crooked table.} The blue curve denotes the stability criterion (\ref{tilt2}), while the red solid line is the (approximate) trajectory of the ball with $\theta(t_1)+2\pi=\theta(t_2)$ indicating two successive times of complete revolutions. The point at which the ball leaves the rim will therefore be the first intersection of this stability criterion and the trajectory. This will necessarily be in the region to the left of the point at which the ball's trajectory is tangent to (\ref{tilt2}), and this is highlighted in the figure as a green solid. Typically a crooked table will only be slightly crooked and hence this region will be close to $\theta=0$ but biased toward the approaching ball. The width of that region depends on  $\dot\theta(t_1)-\dot\theta(t_2)$, which in turn can be determined from (\ref{tilt2}).}
\label{tiltfig}
\end{figure} 

Suppose, now that the table is not perfectly level. This is the situation discussed and exploited by Thorp\cite{eT69}. Without loss of generality (it is only an affine change of co-ordinates for any other orientation) suppose that the table is tilted by an angle $\delta$ such that the origin $\varphi=0$ is the lowest point on the rim.  Just as with the case of a level table, the time which the ball spends in the rim is variable and the time at which it leaves the rim depends on a stability criterion similar to $(\ref{inrim})$. But now that the table is not level, that equilibrium becomes
\begin{eqnarray}
\label{tilt}
r\dot\theta^2&=&g\tan{\left(\alpha+\delta\cos\theta\right)}.
\end{eqnarray} 
If $\delta=0$ then it is clear that the distribution of angular positions for which this condition is first met will be uniform. Suppose instead that $\delta>0$, then there is now a range of critical angular velocities $\dot\theta^2_{\rm crit}\in[\frac gr\tan{\left(\alpha-\delta\right)},\frac gr\tan{\left(\alpha+\delta\right)}]$. Once $\dot\theta^2<{\frac gr\tan{\left(\alpha+\delta\right)}}$ the position at which the ball leaves the rim will be dictated by the point of intersection in $(\theta,\dot\theta)$-space of 
\begin{eqnarray}
\label{tilt2}
\dot\theta &= &\sqrt{\frac gr\tan{\left(\alpha+\delta\cos\theta\right)}}
\end{eqnarray}
and the ball trajectory as a function of $t$ (modulo $2\pi$)
\begin{eqnarray}
\label{dtheta}
\dot\theta(t) & = & \dot\theta(0)+\ddot\theta(0)t \\
\label{theta}
\theta(t) & = & \theta(0)+\dot\theta(0)t+\frac 12 \ddot\theta(0)t^2.
\end{eqnarray}
If the angular velocity of the ball is large enough then the ball will leave the rim at some point on the half circle prior to the low point ($\varphi=0$). Moreover, suppose that in one revolution (i.e. $\theta(t_1)+2\pi=\theta(t_2)$ ), the velocity changes by $\dot\theta(t_1)-\dot\theta(t_2)$. Furthermore, suppose that this is the first revolution during which $\dot\theta^2<{\frac gr\tan{\left(\alpha+\delta\right)}}$ (that is, $\dot\theta(t_1)^2\geq{\frac gr\tan{\left(\alpha+\delta\right)}}$ but $\dot\theta(t_2)^2<{\frac gr\tan{\left(\alpha+\delta\right)}}$). Then, the point at which the ball will leave the rim will (in $(\theta,\dot\theta)$-space) be the intersection of (\ref{tilt2}) and 
\begin{eqnarray}
\label{thetaline}
\dot\theta & = & 
\dot\theta(t_1)-\frac{1}{2\pi}\left(\dot\theta(t_2)-\dot\theta(t_1)\right)\theta.
\end{eqnarray}
The situation is depicted in figure \ref{tiltfig}. One can expect for a tilted roulette wheel, the ball will systematically favour leaving the rim on one half of the wheel. Moreover, to a good approximation, the point at which the ball will leave the rim follows a uniform distribution over significantly less than half the wheel circumference. In this situation, the problem of predicting the final resting place is significantly simplified to the problem of predicting the position of the {\em wheel} at the time the ball leaves the rim. 

We will pursue this particular case no further here. The situation (\ref{tilt}) may be considered as a generalisation of the ideal $\delta=0$ case. This generalisation makes the task of prediction significantly easier, but we will continue to work under the assumption that the casino will be doing its utmost to avoid the problems of an improperly levelled wheel. Moreover, this generalisation is messy, but otherwise uninteresting.  In the next section we consider the problem of implementing a prediction scheme for a perfectly level wheel.

\section{Experimental results}
\label{predict}

In Sec. \ref{model} we introduces the basic mathematical model which we will utilise for the prediction of the trajectory of the ball within the roulette wheel. We ignore (or rather treat as essentially stochastic) the trajectory of the ball after hitting the deflectors --- charting the distribution of final outcome from deflector to individual pocket in the roulette wheel is a tractable probabilistic problem, and one which for which we will sketch a solution later. However, the details are perhaps only of interest to the professional gambler and not to most physicists. Hence, we are reduced to predicting the location of the wheel and the ball when the ball first reaches one of the deflectors. The model described in Sec., \ref{model} is sufficient to achieve this --- provided one has adequate measurements of the physical dimensions of the wheel and all initial positions, velocities and accelerations  (as a further approximation we assume deceleration of both the ball and wheel to be constant over the interval which we predict. 
  
Hence, the problem of prediction is essentially two-fold. First, the various velocities must be estimated accurately. Given these estimates it is a trivial problem to then determine the point at which the ball will intersect with one of the deflectors on the stator. Second, one must then have an estimate of the scatter imposed on the ball by both the deflectors and possible collision with the individual frets. To apply this method {\em in situ}, one has the further complication of estimating the parameters $r$, $r_{\rm defl}$, $r_{\rm rim}$, $\alpha$ and possibly $\delta$ without attracting undue attention. We will ignore this additional complication as it is essentially a problem of data collection and statistical estimation. Rather, we will assume that these quantities can be reliably estimated and restrict our attention to the problem of prediction of the motion.  To estimate the relevant positions, velocities and accelerations $(\theta,\dot\theta,\ddot\theta,\varphi,\dot\varphi,\ddot\varphi)_{t=0}$ (or perhaps just  $(\theta-\varphi,\dot\theta-\dot\varphi,\ddot\theta-\ddot\varphi)_{t=0}$) we employ two distinct techniques. 

In the following subsections we describe these methods. In Sec. \ref{manual} we introduce a manual measurement scheme, and in Sec. \ref{auto} we describe our implementation of a more sophisticated digital system. The purpose of Sec. \ref{manual} is to demonstrate that a rather simple ``clicker'' type of device --- along the lines of that utilized by the Doyne Farmer, Norman Packard and collaborators \cite{tB90} --- can be employed to make sufficiently accurate measurements. Nonetheless, this system is far from optimal: we conduct only limited experiments with this apparatus: sufficient to demonstrate that, in principle, the method could work. In Sec. \ref{auto} we describe a more sophisticated system. This system relies on a digital camera mounted directly above a roulette wheel and is therefore unlikely to be employed in practice (although alternative, more subtle, devices could be imagined). Nonetheless, our aim here is to demonstrate how well this system could work in an optimal environment.

Of course, the degree to which the model in Sec. \ref{model} is able to provide a useful prediction will depend critically on how well the parameters are estimated. Sensitivity analysis shows that predicted outcome (Eq. {\ref{rest}) depends only linearly or quadratically (in the case of physical parameters of the wheel) on our initial estimates. More important however, and more difficult to estimate, is to what extent each of these parameters can be reliably estimated. For this reason we take a strictly experimental approach and show that even with the various imperfections inherent in experimental measurement, and in our model, sufficiently accurate predictions are realizable.

\subsection{A manual implementation}
\label{manual}

\begin{figure}
\begin{center}
\includegraphics[width=0.45\textwidth]{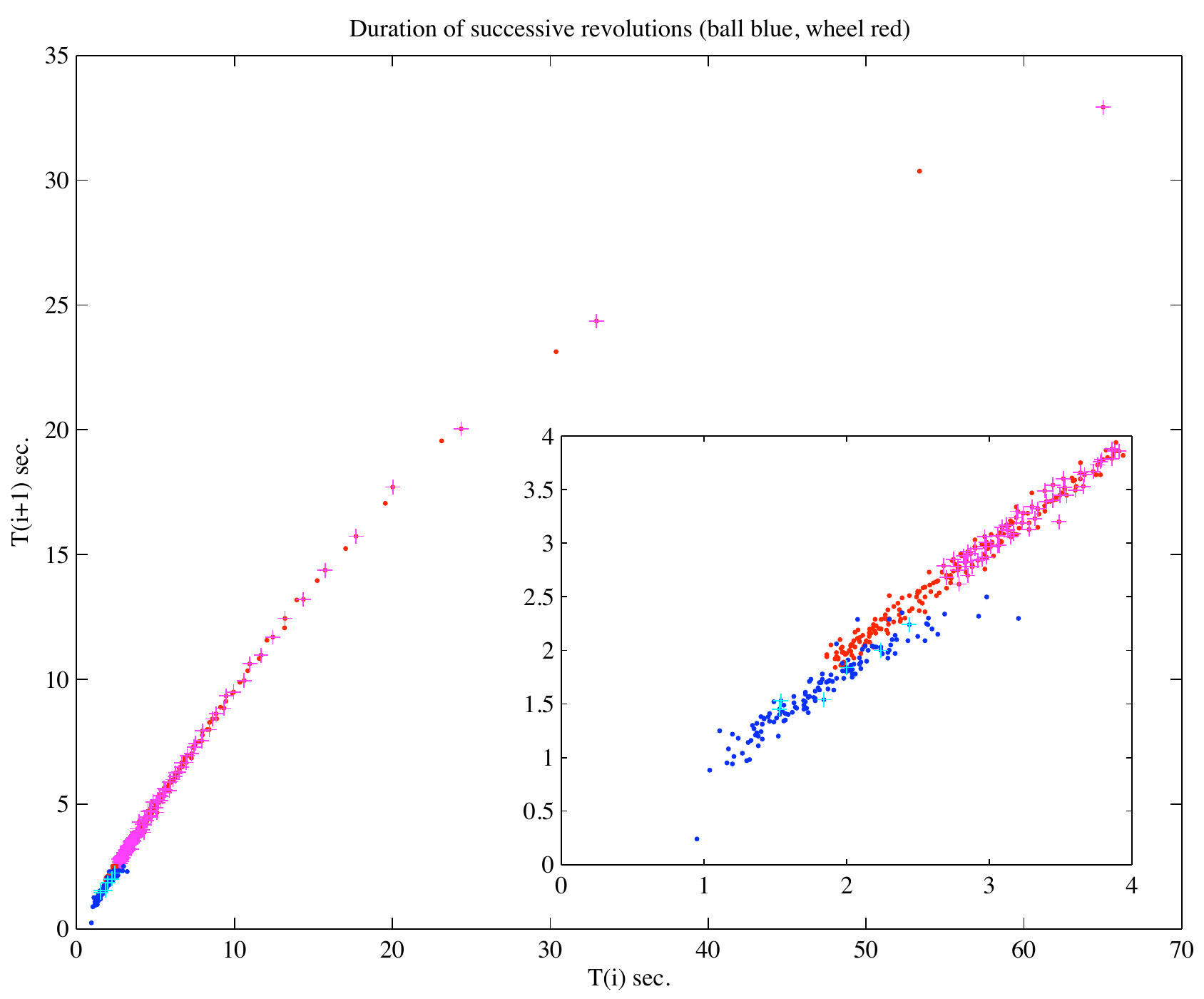}
\end{center}
\caption{{\bf Hand-measurement of ball and wheel velocity for prediction.} From two spins of the wheel, and 20 successive spins of the ball we logged the time (in seconds) $T(i)$ for successive passes past a given point ($T(i)$ against $T(i+1)$). The measurements $T(i)$ and  $T(i+1)$ are the timings of successive revolutions --- direct measurements of the angular velocity observed over one complete rotation. To provide the simplest and most direct indication that handheld measurements of this quantity are accurate, we indicate in this figure a deterministic relationship between these quantities. From this relationship one can determine the angular deceleration.  The red points depict these times for the wheel, the blue points are for the ball. A single trial of both ball and wheel is randomly highlighted with crosses (superimposed). The inset is an enlargement of the detail in the lower left corner. Both the noise and the determinism of this method are evident. In particular, the wheel velocity is relatively easy to calculate and decays slowly, in contrast the ball decays faster and is more difficult to measure. }
\label{manualdata}
\end{figure} 

Our first approach is to simply record the time at which ball and wheel pass a fixed point. This is a simple approach (probably that used in the early attempts to beat the wheels of Las Vegas) and is trivial to implement on a laptop computer, personal digital assistant, embedded system, or even a mobile phone\footnote{Implementation on a ``shoe-computer'' should be relatively straightforward too.}$^,$\cite{fyp1,fyp2}. Our results, depicted in figure \ref{manualdata} illustrate that the measurements, although noisy, are feasible. The noise introduced in this measurement is probably largely due to the lack of physical hand-eye co-ordination of the first author. Figure \ref{manualdata} serves only to demonstrate that, from measurements of successive revolutions $T(i)$ and $T(i+1)$ the relationship between $T(i)$ and $T(i+1)$ can be predicted with a fairly high degree of certainty (over several trials and with several different initial conditions. As expected, the dependence of $T(i+1)$ on $T(i)$ is sub-linear. Hence, derivation of velocity and acceleration from these measurements should be relatively straightforward. Simple experiments with this configuration indicate that it is possible to accurately predict the correct half of the wheel in which the ball will come to rest (statistical results are given in the caption of Fig. \ref{manualdata}).

Using these (admittedly noisy) measurements we were able to successfully predict the half of the wheel in which the ball would stop in $13$ of $22$ trials (random with $p<0.15$), yielding an expected return of $36/18\times 13/22-1=+18\%$. This trial run included predicting the precise location in which the ball landed on three occasions (random with $p<0.02$). Quoted $p$-values are computed against the null hypothesis of a binomial distribution of $n$ trials with probability of success determine by the fraction of the total circumference corresponding to the target range --- i.e. the probability of landing by chance in one of the target pockets.

\subsection{Automated digital image capture}
\label{auto}

Alternatively, we employ a digital camera mounted directly above the wheel to accurately and instantaneously measure the various physical parameters. This second approach is obviously a little more difficult to implement incognito. Here, we are more interested in determining how much of an edge  can be achieved under ideal conditions, rather than the various implementation issues associated with realising this scheme for personal gain. In all our trials we use a regulation casino-grade roulette wheel (a $32$" ``President Revolution'' roulette wheel manufactured by Matsui Gaming Machine Co. Ltd., Tokyo). The wheel has 37 numbered slots (1 to 36 and 0) in the configuration shown in figure \ref{wheel} and has a radius of $820$ mm (spindle to rim). For the purposes of data collection we employ a Prosilica EC650C IEEE-1394 digital camera (1/3" CCD, 659$\times$493 pixels at 90 frames per second). Data collection software was written and coded in C++ using the OpenCV library.
 
The camera provides approximately (slightly less due to issues with data transfer) 90 images per second of the position of the roulette wheel and the ball. Artifacts in the image due to lighting had to be managed and filtered. From the resultant image the position of the wheel was easily determined by locating the only green pocket (``0'') in the wheel, and the position of the ball was located by differencing successive frames (searching for the ball shape or color was not sufficient due to the reflective surface of the wheel and ambient lighting conditions). 

From these time series of Cartesian coordinates for the position of both the wheel (green ``0'' pocket) and ball we computed the centre of rotation and hence derived angular position time series. Polynomial fits to these angular position data (modulo $2\pi$) provided estimates of angular velocity and acceleration (deceleration). From this data we found that, for out apparatus, the acceleration terms where very close to being constant over the observation time period --- and hence modeling the forces acting on the ball as constant provided a reasonable approximation. With these parameters we directly applied the model of Section \ref{model} to predict the point at which the ball came into contact with the deflectors. 

\begin{figure}
\begin{center}
\includegraphics[width=0.45\textwidth]{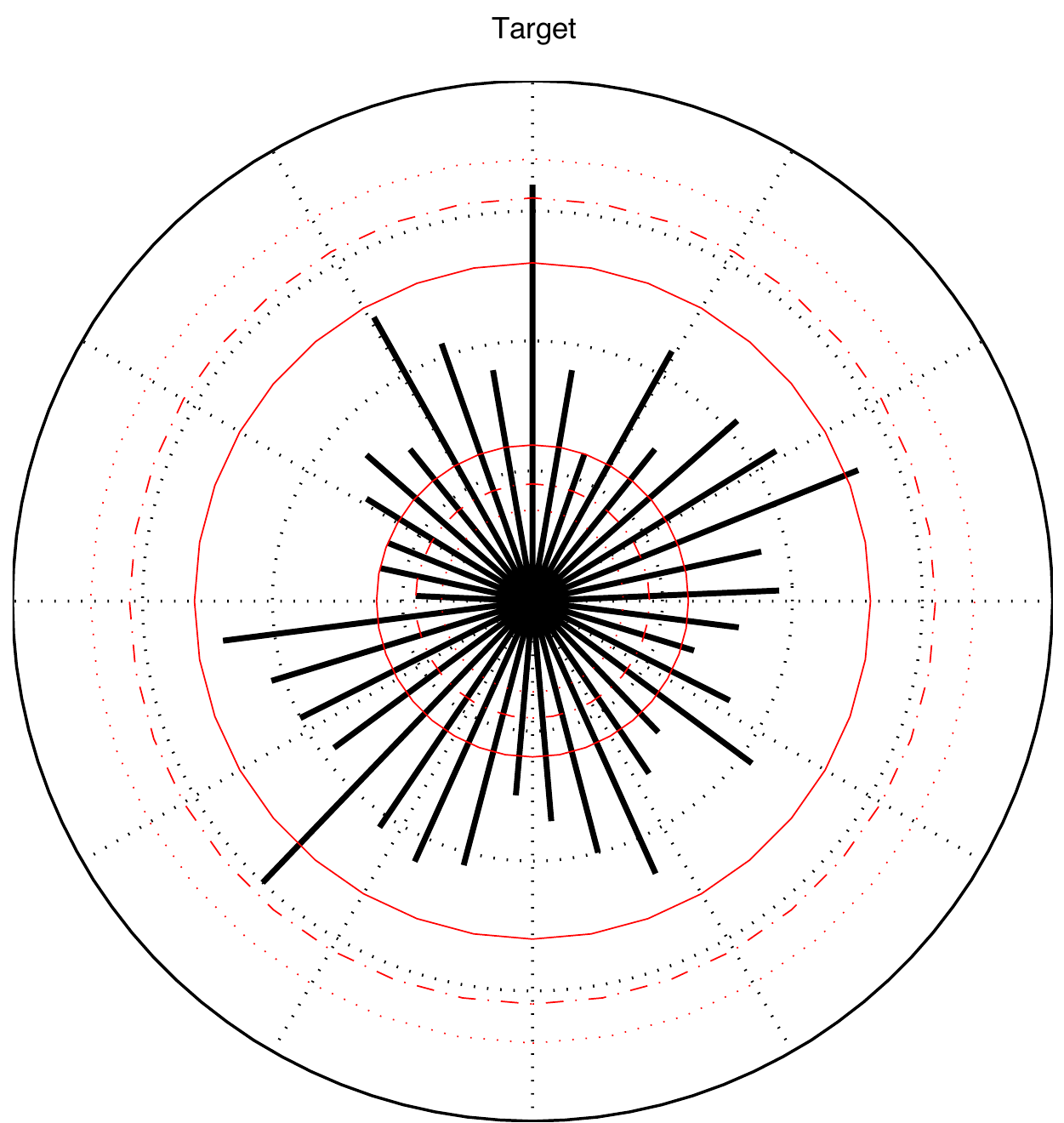}
\end{center}
\caption{{\bf Predicting roulette.} The plot depicts the results of $700$ trials of our automated image recognition software used to predict the outcome of independent spins of a roulette wheel. What we plot here is a histogram in polar coordinates of the difference between the predicted and the actual outcome (the``Target" location, at the 12 o'clock position in this diagram,  indicating that the prediction was correct). The length of each of the 37 black bars denote the frequency with which predicted and actual outcome differed by exactly the corresponding angle. Dot, dot-dashed and solid (red) lines depict the corresponding $99.9\%$, $99\%$ and $90\%$ confidence intervals using the corresponding two-tailed binomial distribution. Motion forward (i.e. ball continues to move in the same direction) is clockwise, motion backwards in anti-clockwise. From the $37$ possible results there are $2$ instances outside the $99\%$ confidence interval. There are $7$ instances outside the $90\%$ confidence interval. }
\label{finalres}
\end{figure} 

Figure \ref{finalres} illustrates the results from $700$ trials of the prediction algorithm on independent rolls of a fair and level roulette wheel. The scatter plot of Fig. \ref{finalres} provides only a crude estimation of variance over the entire region of the wheel for a given prediction. A determined gambler could certainly extend this analysis with a more substantial data set relating to their particular wheel of interest. We only aim to show that certain non-random characteristics in the distribution of resting place will emerge and that these can then be used to further refine prediction.

Nonetheless, several things are clear from Fig. \ref{finalres}. First, for most of the wheel, the probability of the ball landing in a particular pocket --- relative to the predicted destination --- does not differ significantly from chance: observed populations in $30$ of $37$ pockets is within the $90\%$ confidence interval for a random process. Two particular pockets --- the target pocket itself and a pocket approximately one-quarter of the wheel prior to the target pocket --- occur with frequencies higher than and less than (respectively) the expected by chance: outside the $99\%$ confidence interval. Hence, the predicted target pocket is a good indicator of eventual outcome and those pockets immediate prior to the target pocket (which the ball would need to bounce backwards to reach) are less likely. Finally, and rather speculatively, there is a relatively higher chance (although marginally significant) of the ball landing in one of the subsequent pockets --- hence, suggesting that the best strategy may be to bet on the section of the wheel following the actual predicted destination. 

\section{Exploits and counter-measures}
\label{exploit}

The essence of the method presented here is to predict the location of the ball and wheel at the point when the ball will first come into contact with the deflectors. Hence, we only require knowledge of initial conditions of each aspect of the system (or more concisely, their relative positions, velocities and accelerations). In addition to this, certain parameters derived from the physical dimensions of the wheel are required --- these could either be estimated directly, or inferred from observational trajectory data. Finally, we note that while anecdotal evidence suggests that (the height of the) frets plays an important role in the final resting place of the ball, this does not enter into our model of the more deterministic phase of the system dynamics. It will affect the distribution of final resting places --- and hence this is going to depend rather sensitively on a particular wheel.

We would like to draw two simple conclusions from this work. First, deterministic predictions of the outcome of a game of roulette can be made, and can probably be done in situ. Hence, the tales of various exploits in this arena are likely to be based on fact. Second, the margin for profit is quite slim. Minor manipulation with the frictional resistance or level of the wheel and/or the manner in which the croupier actually plays the ball (the force with which the ball is rolled and the effect, for example, of axial spin of the ball) have not been explored here and would likely affect the results significantly. Hence, for the casino the news is mostly good --- minor adjustments will ameliorate the advantage of the physicist-gambler. For the gambler, one can rest assured that the game is on some level predictable and therefore inherently honest.

Of course, the model we have used here is extremely simple. In Strzalko {\em et al.} \cite{jS09} much more sophisticated modeling methodologies have been independently developed and presented. Certainly, since the entire system is a physical dynamical system, computational modeling of the entire system may provide an even greater advantage \cite{jS09}. Nonetheless,  the methods presented in this paper would certainly be within the capabilities of a 1970's ``shoe-computer''.

\section*{Acknowledgement}

The first author would like to thank Marius Gerber for introducing him to the dynamical systems aspects of the game of roulette. Funding for this project, including the roulette wheel, was provided by the Hong Kong Polytechnic University. The labors of final year project students, Yung Chun Ting and Chung Kin Shing, in performing many of the mechanical simulations describe herein is gratefully acknowledged.

\bibliography{../bibliography}
\bibliographystyle{unsrt}

\end{document}